\documentclass[conference]{IEEEtran}
\IEEEoverridecommandlockouts
\usepackage{cite}
\usepackage{amsmath,amssymb,amsfonts}
\usepackage{algorithmic}
\usepackage{graphicx}
\usepackage{textcomp}

\usepackage{hyperref}
\hypersetup{
    colorlinks=true,
    linkcolor=blue,
    filecolor=magenta,      
    urlcolor=cyan,
    pdftitle={Overleaf Example},
    pdfpagemode=FullScreen,
    }

\usepackage{xcolor}
\def\BibTeX{{\rm B\kern-.05em{\sc i\kern-.025em b}\kern-.08em
    T\kern-.1667em\lower.7ex\hbox{E}\kern-.125emX}}

\usepackage[utf8]{inputenc}
\usepackage{colortbl}
\usepackage{array}
\usepackage{caption}
\definecolor{tablepurple}{RGB}{196,179,225} 
\begin{document}

\title{Benchmarking Quantum Convolutional Neural Networks for Signal Classification in Simulated Gamma-Ray Burst Detection
}

\author{\IEEEauthorblockN{Farida Farsian}
\IEEEauthorblockA{\textit{OACT, INAF} \\
Via S. Sofia 78, Catania, Italy \\
farida.farsian@inaf.it}
\and
\IEEEauthorblockN{Nicol\'o Parmiggiani}
\IEEEauthorblockA{\textit{OAS, INAF} \\
Via Pietro Gobetti 93/3, Bologna, Italy\\
nicolo.parmiggiani@inaf.it}
\and
\IEEEauthorblockN{Alessandro Rizzo}
\IEEEauthorblockA{\textit{OACT, INAF} \\
Via S. Sofia 78, Catania, Italy \\
alessandro.rizzo@inaf.it}
\and
\IEEEauthorblockN{Gabriele Panebianco}
\IEEEauthorblockA{\textit{OAS, INAF} \\
Via Pietro Gobetti 93/3, Bologna, Italy \\
gabriele.panebianco@inaf.it}
\and
\IEEEauthorblockN{Andrea Bulgarelli}
\IEEEauthorblockA{\textit{OAS, INAF} \\
Via Pietro Gobetti 93/3, Bologna, Italy \\
andrea.bulgarelli@inaf.it}
\and
\IEEEauthorblockN{Francesco Schillir\'o}
\IEEEauthorblockA{\textit{OACT, INAF} \\
Via S. Sofia 78, Catania, Italy \\
francesco.schilliro@inaf.it}
\and
\IEEEauthorblockN{Carlo Burigana}
\IEEEauthorblockA{\textit{IRA, INAF} \\
Via Pietro Gobetti 101, Bologna, Italy \\
carlo.burigana@inaf.it}
\and
\IEEEauthorblockN{Vincenzo Cardone}
\IEEEauthorblockA{\textit{OAR, INAF} \\
Via Frascati 33, Monteporzio Catone, Italy \\
vincenzo.cardone@inaf.it}
\and
\IEEEauthorblockN{Luca Cappelli}
\IEEEauthorblockA{\textit{OATs, INAF} \\
Via Tiepolo 11, Trieste, Italy \\
luca.cappelli@inaf.it}
\and
\IEEEauthorblockN{Massimo Meneghetti}
\IEEEauthorblockA{\textit{OAS, INAF} \\
Via Pietro Gobetti 93/3, Bologna, Italy \\
massimo.meneghetti@inaf.it}
\and
\IEEEauthorblockN{Giuseppe Murante}
\IEEEauthorblockA{\textit{OATs, INAF} \\
Via Tiepolo 11, Trieste, Italy \\
giuseppe.murante@inaf.it}
\and
\IEEEauthorblockN{Giuseppe Sarracino}
\IEEEauthorblockA{\textit{OACN, INAF} \\
Via Moiariello 16, Napoli, Italy \\
giuseppe.sarracino@inaf.it}
\and
\IEEEauthorblockN{Roberto Scaramella}
\IEEEauthorblockA{\textit{OAR, INAF} \\
Via Frascati 33, Monteporzio Catone, Italy \\
roberto.scaramella@inaf.it}
\and
\IEEEauthorblockN{Vincenzo Testa}
\IEEEauthorblockA{\textit{OAR, INAF} \\
Via Frascati 33, Monteporzio Catone, Italy \\
vincenzo.testa@inaf.it}
\and
\IEEEauthorblockN{Tiziana Trombetti}
\IEEEauthorblockA{\textit{IRA, INAF} \\
Via Pietro Gobetti 101, Bologna, Italy \\
tiziana.trombetti@inaf.it}
}

\maketitle

\begin{abstract}
This study evaluates the use of Quantum Convolutional Neural Networks (QCNNs) for identifying signals resembling Gamma-Ray Bursts (GRBs) within simulated astrophysical datasets in the form of light curves. The task addressed here focuses on distinguishing GRB-like signals from background noise in simulated Cherenkov Telescope Array Observatory (CTAO) data, the next-generation astrophysical observatory for very high-energy gamma-ray science.  
QCNNs, a quantum counterpart of classical Convolutional Neural Networks (CNNs), leverage quantum principles to process and analyze high-dimensional data efficiently. We implemented a hybrid quantum-classical machine learning technique using the Qiskit framework, with the QCNNs trained on a quantum simulator. Several QCNN architectures were tested, employing different encoding methods such as Data Reuploading and Amplitude encoding.  
Key findings include that QCNNs achieved accuracy comparable to classical CNNs, often surpassing 90\%, while using fewer parameters, potentially leading to more efficient models in terms of computational resources. A benchmark study further examined how hyperparameters like the number of qubits and encoding methods affected performance, with more qubits and advanced encoding methods generally enhancing accuracy but increasing complexity.
QCNNs showed robust performance on time-series datasets, successfully detecting GRB signals with high precision. The research is a pioneering effort in applying QCNNs to astrophysics, offering insights into their potential and limitations. This work sets the stage for future investigations to fully realize the advantages of QCNNs in astrophysical data analysis.
\end{abstract}


\section{Introduction}
In the last decade the advancement of quantum computing hardware and its availability to public, prepared the playground to investigate the potential of quantum computing in different scientific cases. Quantum computers offers an incredible potential in speed up and performance on calculation prime numbers \cite{365700}, simulating quantum systems \cite{Lloyd1996UniversalQS}, or solving linear systems of equations \cite{Harrow_2009}.
This rapid advancements in quantum computing have opened new avenues for tackling complex computational problems across various scientific disciplines. Among the most promising applications is Quantum Machine Learning (QML), which merges the computational power of quantum mechanics with the established frameworks of classical machine learning \cite{Zeguendry:2023pbk}. QML offers powerful tools to explore the frontiers of science, particularly in Astrophysics.

In the field of astrophysics, detecting and analyzing transient astrophysical events, such as Gamma-Ray Bursts (GRBs), presents a significant challenge. GRBs are among the most energetic and enigmatic phenomena in the universe, requiring sophisticated techniques to capture, process, and interpret their fleeting signals. 

Quantum Convolutional Neural Networks (QCNNs) represent an innovative hybrid quantum-classical machine learning framework inspired by the hierarchical structure of classical convolutional neural networks (CNNs). Unlike classical CNNs, QCNNs utilize parameterized quantum circuits to exploit the advantages of quantum mechanics, enabling efficient representation and processing of complex data. The QCNN architecture combines translationally invariant unitary operations and pooling mechanisms that reduce the system's degrees of freedom, allowing for scalable learning with \( O(\log(N)) \) variational parameters for input sizes of \( N \) qubits. This makes QCNNs particularly well-suited for near-term quantum devices and applications requiring both computational efficiency and robust pattern recognition \cite{2019NatPh..15.1273C}.

Recent developments in QML have demonstrated its potential in various areas of physics. In high-energy physics, for instance, QML algorithms have been proposed to address computational challenges associated with large-scale data analysis, as highlighted in a review of quantum approaches to particle physics applications \cite{Guan:2020bdl}.  
Similarly, astrophysics and cosmology have seen emerging applications of QML. For instance, quantum-enhanced support vector machines have been successfully applied to galaxy classification, demonstrating the ability to process high-dimensional astronomical datasets \cite{Hassanshahi:2023xgv}. In another study, QML was utilized in radio astronomy, where a quantum neural network was developed for pulsar classification, leveraging quantum-specific encoding methods to achieve comparable accuracy to classical techniques \cite{Kordzanganeh:2021azm}.
Despite these advances, the potential of QCNNs in astrophysics remains largely unexplored. This paper aims to fill this gap by systematically studying the capabilities of QCNNs for analyzing astrophysical data, specifically in the context of Gamma-Ray Burst (GRB) detection.

Despite all the potentials, QML faces key challenges, including the trainability of quantum models, which can be hindered by barren plateaus—regions of exponentially flat loss landscapes—as well as noise in quantum hardware that can corrupt results and complicate optimization processes. Additionally, embedding classical data into quantum states remains a non-trivial task, requiring efficient encoding methods that preserve the computational advantages of the quantum system \cite{Cerezo:2022nvi}.
Therefore, it is crucial to systematically benchmark different quantum architectures to identify the most effective models for specific applications. Benchmarking these different architectures, including the data encoding, is essential to understanding their relative strengths and weaknesses in real-world tasks. In the same line, \cite{Bowles:2024fvp} have done a extensive work on benchmarking different QML architecture for the binary classification task.

In the context of GRB detection with CNN, there has been previous works to detect GRBs using classical deep learning methods \cite{Parmiggiani:2023kum, Parmiggiani_2024, 2023ApJ...945..106P}. These approaches utilize CNNs to process time-series data from GRB signals, aiming to classify and analyze transient astrophysical phenomena. The CNN models used in these works show an advantage respect to standard method used in AGILE space mission \cite{AGILE:2008nyq} data analysis pipeline and can recognize more GRBs. Classical models have demonstrated success in automating GRB detection with very high accuracy. In this work we would like to explore the potential of quantum machine learning techniques for advancing GRB studies.
It is important to note that in this study, the task of GRB detection is framed as a binary classification problem, where the goal is to distinguish GRB signals from background noise. The performance of our method is evaluated using accuracy, which, as commonly defined in the machine learning context, serves as a measure of the model's correctness in correctly classifying the GRB signals and background noise.

In our previous study \cite{Rizzo:2024wse}, we investigated the feasibility of employing QCNNs for GRB detection using datasets from the AGILE space mission \cite{AGILE:2008nyq}. This initial work evaluated the QCNN's performance in processing both time-series data and sky maps, utilizing the \textit{Qiskit} and \textit{Pennylane} \cite{Bergholm:2018cyq} quantum computing platforms. The results demonstrated the potential of QCNNs for this application, indicating promising capabilities in GRB detection tasks.  

Building on these findings, the present study aims to thoroughly assess the QCNN's strengths and limitations in this context. Specifically, we explore its full potential by systematically evaluating its performance across various configurations and parameters, with the goal of identifying both the robust aspects of the model and areas that require optimization or improvement. This comprehensive evaluation seeks to advance our understanding of QCNNs' applicability to real-world astrophysical signal detection challenges.

The structure of this paper is as follows: Section II provides an overview of QML, including its foundational principles and integration with classical methods. Section III describes the dataset used for QCNN training and testing, focusing on simulated GRB signals and background noise. Section IV outlines the QCNN architecture, detailing the variational circuits, data encoding methods, and benchmarking configurations. Section V presents the experimental results, including performance comparisons with classical CNNs, the impact of hyperparameters, and the generalization capability with limited training data. Finally, Section VI summarizes the findings and explores future directions in QML for astrophysical applications.


\section{Quantum Machine Learning}
QML is an emerging interdisciplinary field that combines the principles of quantum computing with the methodologies of classical machine learning. As quantum computing leverages the unique properties of quantum mechanics such as superposition, entanglement, and quantum parallelism, it offers the potential to significantly accelerate computational tasks that are traditionally challenging for classical computers.

QML algorithms are designed to operate on quantum data or use quantum circuits to process classical data in ways that classical algorithms cannot easily replicate. These quantum algorithms are often expressed through parameterized quantum circuits (PQCs), which can be optimized similarly to classical neural networks. PQCs consist of parameterized quantum gates, which are quantum operations whose behavior is controlled by tunable parameters. Examples of such gates include rotation gates $R_x(\theta)$, $R_y(\theta)$, and $R_z(\theta)$, where the angle $\theta$ serves as the parameter. By combining these gates into a circuit, PQCs provide a flexible framework for representing quantum states \cite{Crooks:2019zff}.

PQCs bridge quantum and classical computing: the quantum computer estimates a quantity, while the classical computer optimizes the parameters. This process iterates, continually refining the quantum state \cite{Mitarai_2018}. By tuning these parameters using classical optimization methods, QML aims to find quantum states that provide better solutions to specific problems, ranging from classification and clustering to more complex pattern recognition tasks.

The primary advantage of QML lies in its ability to potentially solve certain problems with exponentially fewer resources compared to classical approaches. However, the field is still in its infancy, with ongoing research focused on developing practical algorithms, understanding their computational advantages, and exploring real-world applications in various domains such as chemistry, finance, and, as this paper explores, astrophysical signal detection.

\subsection{Variational Quantum Algorithm}

One of the most widely used algorithms in quantum machine learning (QML) is the Variational Quantum Algorithm (VQA), a hybrid quantum-classical approach designed to work effectively on Noisy, Intermediate-Scale Quantum (NISQ) devices. A VQA consists of a parameterized quantum circuit (variational ansatz) whose parameters are optimized using a classical optimization algorithm to minimize a cost function. The cost function encodes the problem to be solved and is evaluated using measurements from the quantum circuit.
VQAs have demonstrated versatility across tasks such as quantum system simulation, combinatorial optimization, and eigenvalue estimation. Their success is attributed to their ability to balance quantum resources—such as entanglement and superposition—with classical computational power to mitigate errors and noise. By leveraging problem-specific ansatz designs and advanced optimization techniques, VQAs can achieve efficient parameterization, reducing the quantum circuit depth and the number of measurements required for convergence \cite{Cerezo_2021}. 

\subsection{Data Encoding methods}
Quantum data encoding is the process of mapping classical data into a quantum state to leverage quantum computational advantages. This step is foundational in QML as it determines how efficiently and accurately classical data can be represented and processed within a quantum system. Depending on the nature of the data and the quantum algorithm, different encoding strategies are used, each with trade-offs in terms of resource requirements, scalability, and expressivity. Here we explain two well-known data encoding methods used in QML and in this work:

\textbf{Amplitude Encoding}
maps classical data into the amplitudes of a quantum state. For a normalized classical vector \(\vec{x} = (x_1, x_2, \ldots, x_{2^n})\), the corresponding quantum state is:
\[
|\psi\rangle = \sum_{i=0}^{2^n-1} x_i |i\rangle.
\]
This approach is highly efficient in terms of qubit usage, as it allows \(2^n\) data points to be encoded using only \(n\) qubits. However, preparing such states can be computationally expensive, as it often involves complex unitary transformations, especially for large datasets.


\textbf{Data Re-uploading} method extends the encoding capabilities by repeatedly embedding classical data into the quantum circuit at multiple layers of the quantum computation process. This iterative re-encoding is achieved by applying data-dependent rotations (e.g., \(R_x\), \(R_y\), \(R_z\)) or parameterized gates in multiple rounds. For example, a single data point \(x\) can influence gates across multiple layers of a parameterized quantum circuit:
\[U(\vec{\theta}, \vec{x}) = U_L(\theta_L, f(x)) \ldots U_1(\theta_1, f(x)).\]

where: \(U\) represents the overall unitary transformation applied by the quantum circuit, which evolves the quantum state based on the input data and tunable parameters. \(L\) is the number of layers in the circuit, indicating how many times the data \(x\) is re-embedded. \(\vec{\theta} = \{\theta_1, \theta_2, \ldots, \theta_L\}\) represents the set of trainable parameters that adjust the behavior of the parameterized gates at each layer. \(f(x)\) is a feature mapping function that transforms the input data \(x\) into a form compatible with the quantum gates (e.g., scaling or encoding as rotation angles). \(U_i(\theta_i, f(x))\) corresponds to the parameterized quantum gates applied in the \(i\)-th layer, which depend both on the trainable parameter \(\theta_i\) and the feature-transformed data \(f(x)\).
This method is particularly powerful in scenarios where qubit resources are limited, as it allows a smaller number of qubits to process complex or high-dimensional data by leveraging the circuit depth. It enhances expressivity and has shown success in tasks such as quantum classification and regression. However, it increases the circuit complexity, which might affect execution fidelity on near-term quantum hardware\cite{P_rez_Salinas_2020}.

Each of these methods has specific advantages and is chosen based on the problem's requirements, the quantum algorithm being implemented, and the available quantum hardware. In the context of QCNNs, the encoding strategy often plays a crucial role in determining the network's efficiency and performance.
In this work based on our need (encoding more classical data as much as possible) we have checked the performance of amplitude and data-reuploading methods.

\section{Dataset}

GRBs are extremely energetic explosions that occur in distant galaxies, emitting intense bursts of gamma rays, the most energetic form of light.
They are typically classified into two types: short-duration GRBs, lasting less than 2 seconds, likely caused by the merger of neutron stars, and long-duration GRBs, lasting over 2 seconds, usually associated with the collapse of massive stars into black holes \cite{2009A&A...496..585G}.
GRBs are among the brightest and most powerful events in the universe, often followed by an afterglow that can be observed in other wavelengths like X-rays, visible light, and radio waves. They serve as probes for studying the early universe and testing fundamental physics, including general relativity and Lorentz invariance violationsnables advancements in multimessenger astronomy by linking electromagnetic signals with gravitational waves and neutrinos \cite{Fynbo:2006du, 2009astro2010S.284S}. In this work, we focus on studying long-duration GRBs detection. 

Over the last decade, numerous satellites and telescopes have significantly advanced the observation of gamma rays and GRBs, enabling groundbreaking discoveries. The Fermi Gamma-ray Space Telescope (launched in 2008) and its Large Area Telescope (LAT) and Gamma-ray Burst Monitor (GBM) have been instrumental in GRB detection and spectral analysis \cite{Thompson:2022ufx, 2020ApJ...893...46V, 2021ApJ...913...60P}. NASA's Swift Observatory continues to play a crucial role in the localization and multi-wavelength follow-up of GRBs \cite{SwiftScience:2004ykd}. The AGILE satellite, launched by the Italian Space Agency, focuses on transient gamma-ray phenomena, including GRBs \cite{AGILE:2008nyq, 2022ApJ...925..152U}. 
The Cherenkov Telescope Array Observatory (CTAO), the next generation of ground-based observatories for high and very-high energy science, will enhance gamma-ray astronomy with over a hundred highly sensitive, fast-reacting Cherenkov telescopes. The facility will be equipped with real-time analysis software that automatically generates science alerts and analyzes ongoing observational data in real-time \cite{CTAConsortium:2017dvg, doi:10.1142/10986}.

\subsection{The simulation specification}
For our simulations, we utilized the \textit{gammapy} framework, a widely used Python library for high-level gamma-ray astronomy analysis \cite{gammapy:2023}. Gammapy provides robust tools for simulating, analyzing, and modeling gamma-ray data in formats compatible with Gamma Astronomy Data Format (GADF)\cite{universe7100374}. It also includes support for instrument response functions (IRFs) and event processing, making it suitable for detailed astrophysical studies \cite{gammapy:2024}. 

This simulation approach allowed us to generate a diverse dataset, capturing realistic transient source variability and incorporating features critical for training and evaluating our QCNN. We generated 600 simulated light curves in total from which 440 and 160 are used for training and test set accordingly. Our simulated dataset is balanced, meaning half of the dataset includes the GRB signal and the other half contains only background noise.
The simulation details is as follows:

\textbf{Source Model}:  
The simulated source was a transient point-like gamma-ray emitter located at a 0.4° offset from the center of the field of view, which is a standard configuration for very high energies observations. The spectral model was defined as a power law with an index of \(2.25\), characteristic of Crab-like sources.
It is important to note that no prompt TeV gamma-ray emission has been conclusively observed to date. Consequently, a "typical" spectrum was adopted, with the Crab serving as a standard candle for TeV astronomy. This work does not aim to simulate a realistic GRB prompt emission spectrum, as population studies for GRBs at TeV energies are ongoing \cite{2022icrc.confE.998P}. Developing a fully realistic spectrum is a complex task and lies beyond the scope of this study.
The default power-law amplitude was set to \(3 \times 10^{-10} \, \mathrm{cm^{-2} \, s^{-1} \, TeV^{-1}}\), calibrated to the chosen IRFs. To introduce variability, the amplitude was scaled by a random factor uniformly sampled between \(0.1\) and \(3.0\).

\textbf{Temporal Model}: 
The temporal evolution of the source followed a Gaussian pulse model.
This model is motivated by the first-order approximation of real prompt light curves observed by the Fermi GBM at MeV energies. The prompt emission, usually, is highly irregular, made up by one or more pulses, so to simplify the task, we are using a single smooth pulse model.
The pulse's duration and peak position along the time axis were randomly selected for each simulation. Detecting a GRB signal is important to capture the complete shape of the light curve; therefore, to ensure the inclusion of both the rise and the decline phases of the signal, the Gaussian peak was constrained to avoid the first and last \(200 \, \mathrm{s}\) of the time axis.

\textbf{Simulated Data Products}:  
The simulations generated event lists, which include individual photon events recorded over time. These event lists were subsequently binned to produce counts maps, necessary for downstream analysis. Here we considered 1200 seconds as the duration of our event list which can be binned differently.

\textbf{Instrument Response Functions (IRF)}:
The simulated dataset was generated using the Prod5 IRFs, corresponding to the four Large-Sized Telescopes (LSTs) configuration of the CTAO, with a zenith angle of 20°. The LSTs, with their wide field of view and low-energy detection threshold, are particularly well-suited for observing GRBs, as these features enable the efficient detection of transient, high-energy phenomena across large areas of the sky. These IRFs provide high-precision modeling for gamma-ray observations under standard CTAO performance conditions
\cite{cherenkov_telescope_array_observatory_2021_5499840}.

\begin{figure}[htbp]
\centerline{\includegraphics[width=\columnwidth]{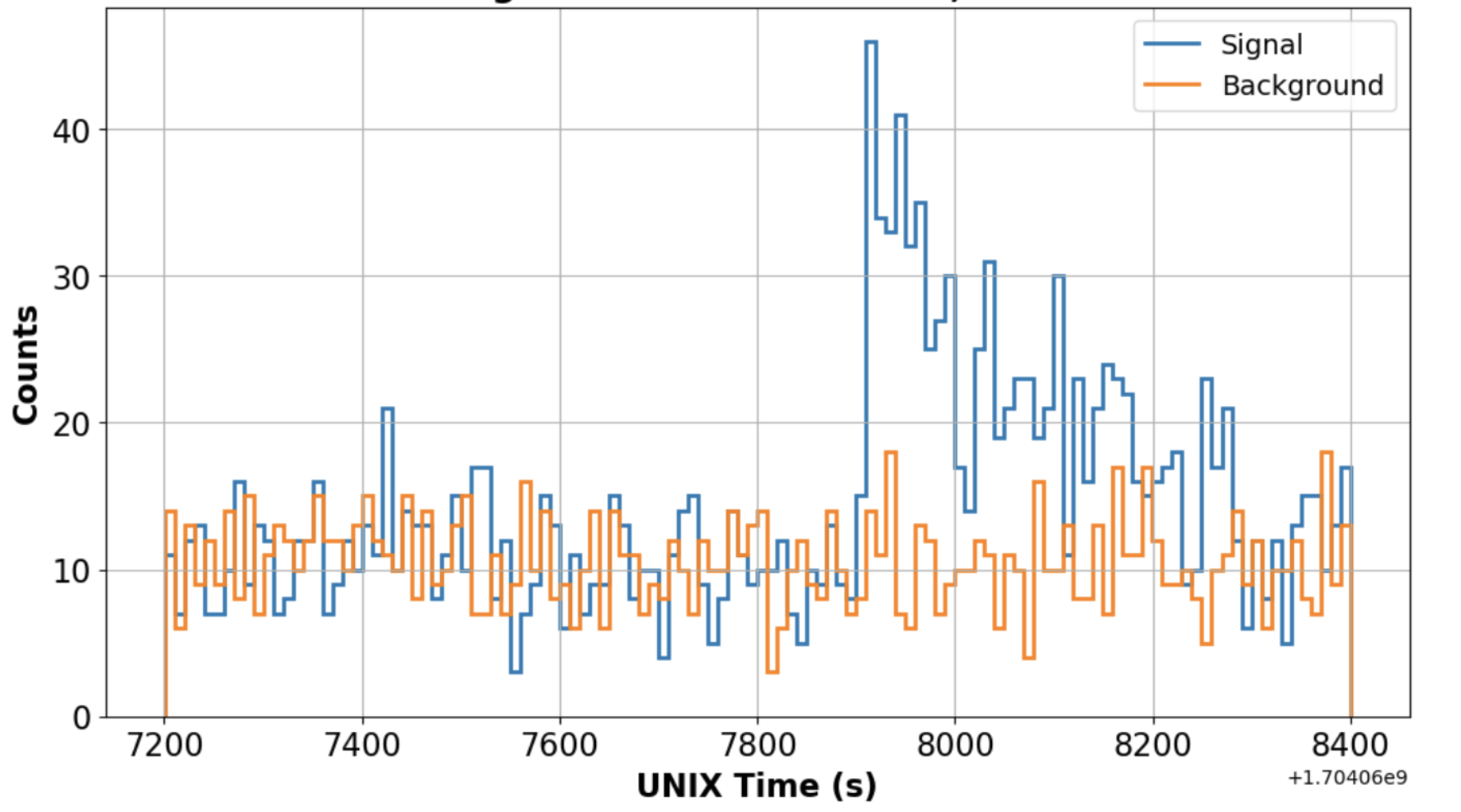}}
\caption{A sample of Simulated Light curves used in training and test set. Y axis show the photon count while the x axis is the time. Blue curve shows the GRB signal with underlying background while orange shows only the background noise. This curve was sampled with bin=10s}
\label{lc_sample}
\end{figure}

Figure \ref{lc_sample} presents a simulated light curve of a GRB. The x-axis represents time, binned into 10-second intervals, while the y-axis corresponds to the photon count in each bin.
The plot effectively demonstrates the GRB event through a distinct peak in the photon count for the signal with underlying background (blue line), which contrasts with the relatively steady background noise (orange line). This specific example highlights a GRB event where the peak is prominent and easily distinguishable.
This light curve serves as a representative sample of the dataset used to train the QCNN. Such simulations are integral to preparing the network to distinguish between signal and background under varying conditions.
While the GRB in this case is evident, the task of identifying GRB events can become considerably more challenging when the signal peak overlaps with or is obscured by background noise. Variability in the occurrence time, intensity, and amplitude of photon counts further complicates detection and classification tasks.
The simulated GRBs in the dataset encompass a range of intensities and amplitudes, mimicking real-world diversity. This variability is crucial for ensuring the robustness of the trained network, enabling it to generalize well to unseen data.

\section{Architecture}
In this section, we detail the implemented architecture for the QCNN as well as the CNN utilized for benchmarking. We also outline the training procedures and hyperparameter configurations employed in our study. By conducting a comprehensive evaluation of different QCNN configurations, including variations in data encoding, qubit count, and network depth, we aim to provide a robust comparison between the quantum and classical approaches.
It is important to note that in the context of our work, the binary classification task is designed to distinguish between GRB signals and background noise.

\subsection{Quantum Convolutional Neural Network}
In this study, we utilized hybrid quantum-classical machine learning, implementing our Quantum Neural Network using Parametrized Quantum Circuits. We adapt the quantum circuit presented in \cite{Sim:2019yyv} according to our needs. We explored the performance of our QCNN using \textit{Qiskit}\footnote{https://www.ibm.com/quantum/qiskit} library.
Qiskit is an open-source quantum computing framework developed by IBM that provides a comprehensive set of tools for building, simulating, and executing quantum circuits on both simulators and real quantum hardware. Designed for accessibility and flexibility, Qiskit supports a range of quantum algorithms, including those for machine learning, and optimization, making it a versatile platform for both research and practical applications in quantum computing.

\begin{figure}[htbp]
\centerline{\includegraphics[width=\columnwidth]{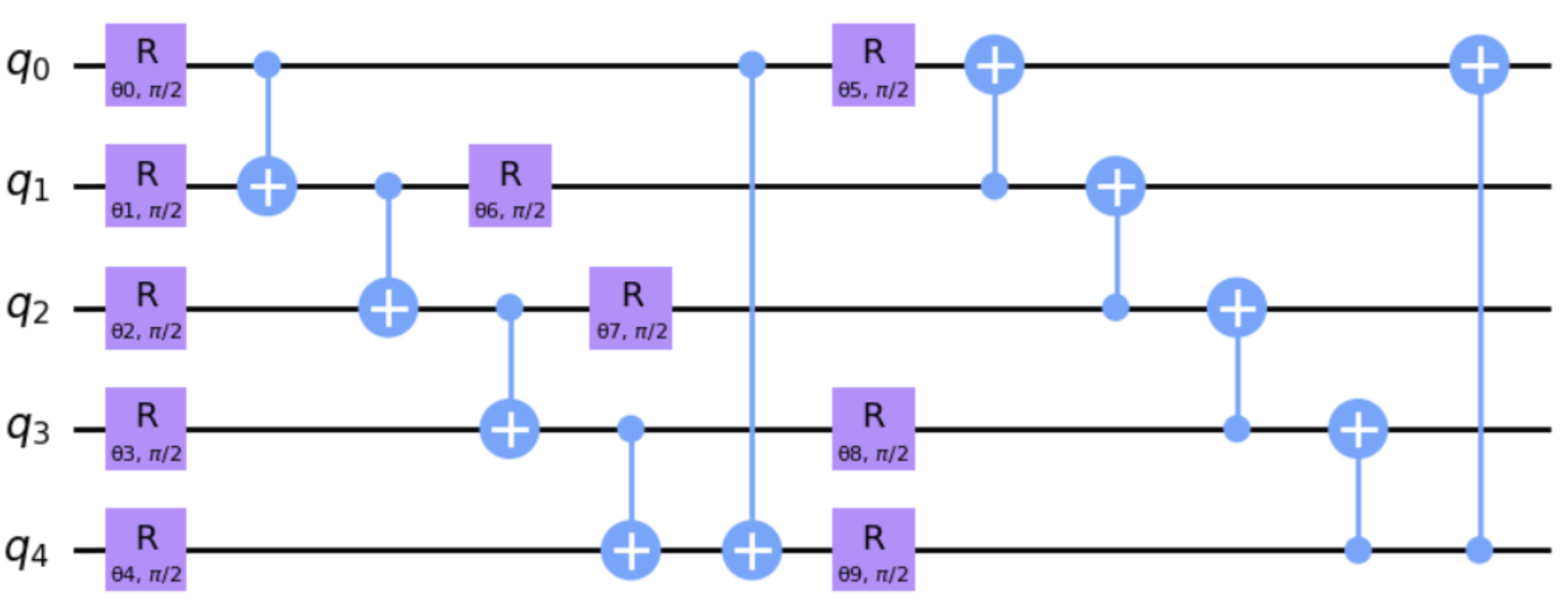}}
\caption{Example of the used quantum circuit in QCNN architecture with 5 qubits, the purple with "R" shows the rotation gates while the blue and "+" sign are CNOT gates demonstration the entanglement operation.}
\label{arch-qcnn}
\end{figure}

The feature map shown in the Figure \ref{arch-qcnn} is an integral component of the QCNN architecture. This feature map architecture uses parameterized single-qubit rotation gates to encode information at the qubit level. Controlled operations are then applied to introduce entanglement between qubits, enabling the quantum representation to capture correlations across qubits. This feature map includes following operations:
\begin{itemize}
    \item Single-Qubit Rotations:  
   Each qubit (\(q_0, q_1, \dots, q_4\)) is subjected to a parameterized single-qubit rotation gate \(R(\theta, \pi/2)\), where \(\theta_i\) are trainable parameters. These rotations initialize the qubits with data-encoded angles, effectively embedding classical input information into the quantum state.
   \item Entanglement Structure:  
   Following the rotation gates, a series of controlled operations (depicted as "+" symbols connected by lines) are applied to create entanglement between neighboring qubits. These operations ensure that the information encoded in one qubit is correlated with others, a key requirement for capturing global patterns and dependencies in the input data.
   \item Layered Connectivity: 
   The entanglement is progressively built across multiple layers, with later connections involving more distant qubits. This hierarchical structure mirrors the pooling or filtering operations in classical CNNs, where localized features are aggregated into more global representations.
\end{itemize}
This feature map plays a critical role in the overall performance of the QCNN, as it determines the quality of the initial quantum representation of the data by embedding data in a way that preserves both individual qubit-level information and inter-qubit correlations. The inclusion of parameterized gates also allows for optimization during training, enabling the network to adaptively learn an optimal embedding for the input data. The use of entanglement ensures that the quantum system leverages its intrinsic advantages over classical systems, particularly in representing high-dimensional and complex data.

\subsection{Classical CNN}
The classical CNN employed in this study is a minimalistic architecture specifically designed to align with the constraints and capabilities of the Quantum QCNN, ensuring a meaningful comparison. The network consists of an input layer followed by a single 1D convolutional layer (Conv1D) with 2 filters and a kernel size of 3, chosen to match the quantum model in terms of complexity and number of trainable parameters. Batch normalization is applied after the convolutional layer to stabilize and accelerate training, followed by a ReLU activation function to introduce non-linearity.  
The feature maps are then passed through a global average pooling (GAP) layer to reduce dimensionality, followed by a flattening operation to prepare the data for dense layers. The dense layer is activated with a ReLU activation function, followed by the output layer, which has as many neurons as the number of classes, using a softmax activation function. The model is trained using a categorical cross-entropy loss function, optimized with the Adam optimizer.
This simplified design was chosen to minimize the number of trainable parameters, ensuring that the comparison focuses on the intrinsic capabilities of QCNN and CNN rather than the complexity of the architecture. By keeping the classical CNN architecture as simple as possible, the study avoids introducing biases that could arise from highly parameterized models.

\subsection{Benchmarking parameters}
To analyze the network performance under varying configurations, we conducted a benchmarking study focusing on the following hyperparameters: the number of qubits, the number of data reuploading layers, the type of data encoding, and the size of the training dataset. We discuss these parameters variation effect in the result section.

\subsection{Training}
\begin{figure}[htbp]
\centerline{\includegraphics[width=0.9\columnwidth]{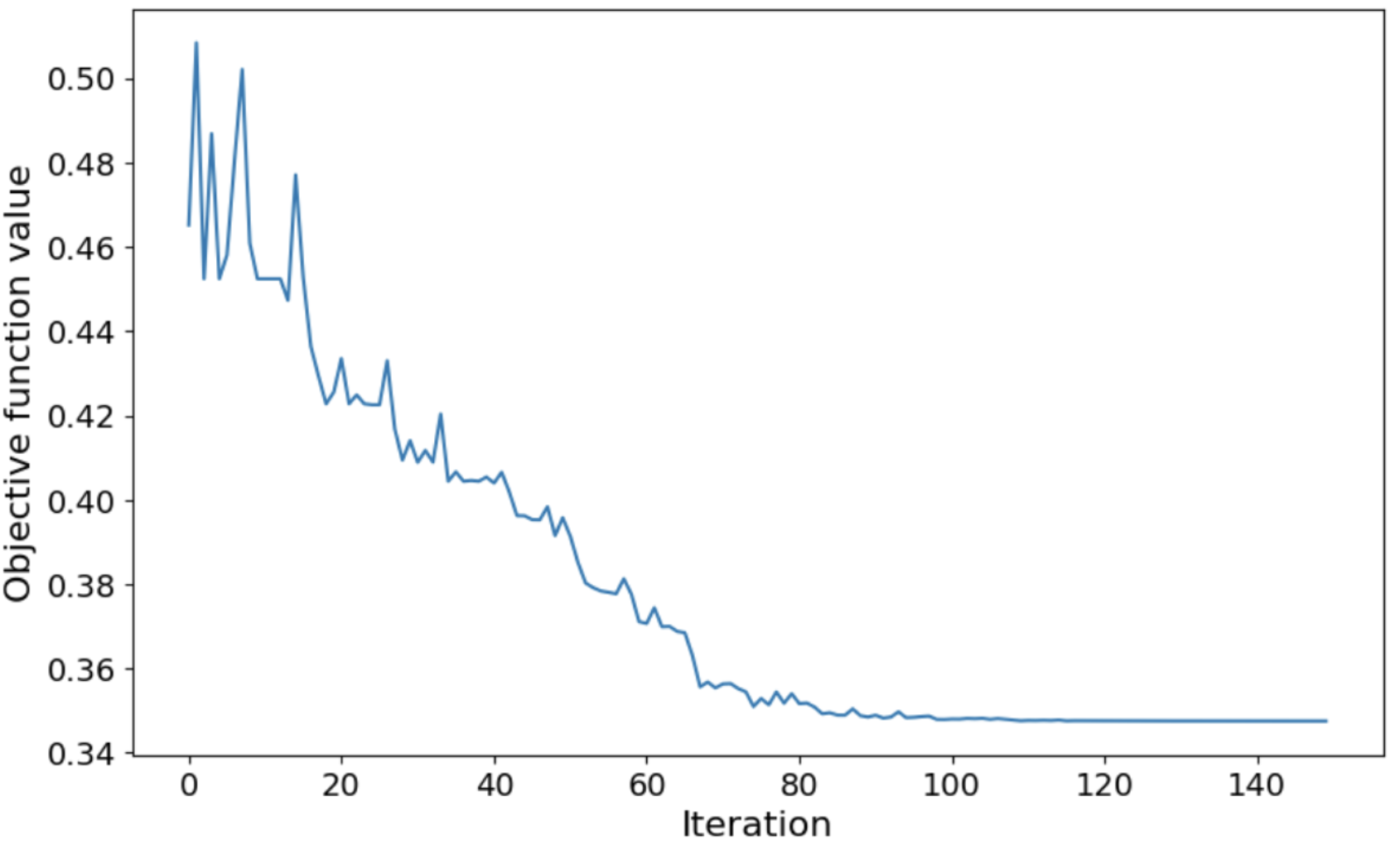}}
\caption{Objectuve function value against iteration during the training the QCNN}
\label{learning_curve}
\end{figure}

Figure \ref{learning_curve} illustrates the convergence of the objective function during the training of the QCNN over 150 iterations. For optimization, we used the COBYLA (Constrained Optimization BY Linear Approximations) algorithm, with a custom callback function to monitor the training process.  QCNN was implemented using the Qiskit SamplerQNN framework and trained on the Aer simulator, which efficiently handles NISQ devices. The initial weights for the variational circuit were initialized randomly within a small range to ensure a well-conditioned starting point for optimization.
The convergence curve shows a general downward trend, indicating that the objective function consistently decreases as the optimization progresses. However, there are fluctuations in the early iterations, due to the randomized initialization of parameters and the non-convex nature of the optimization landscape. After approximately 60 iterations, the curve stabilizes, suggesting that the QCNN has reached a region near the optimal solution. These patterns highlight the importance of a well-tuned optimization process to achieve effective convergence.

Instead, for the classical CNN, we used a batch size of 16 and a variable learning rate when the minimum learning rate is set to \(0.00001\). To prevent overfitting and reduce computational cost, the \textit{early stopping} method was employed, which halted training once the validation loss stopped improving. This method compares the loss value in each epoch, and if the loss starts to increase, which is an indication of overfitting, stops the training. The classical CNN converged in fewer than 400 epochs, demonstrating efficient training under these conditions.  

\section{Results}

In this section, we present the evaluation of the benchmarking results for the implemented models, focusing on the effects of hyperparameters, training sample size, computational time, and model complexity.

\subsection{The General Trend}

The performance evaluation of the QCNN demonstrates that it achieves an accuracy of 97.5\% (as detailed in Table \ref{tab:Num_qubits}), which is comparable to the 97.35\% accuracy achieved by the classical CNN. However, this level of accuracy is attained using fewer parameters, highlighting the potential of QCNNs for parameter-efficient modeling. This efficiency suggests that QCNNs leverage the inherent properties of quantum systems, such as superposition and entanglement, to represent and process data more compactly than classical architectures.


\subsection{The Qubit number effect}
The results indicated in Table \ref{tab:Num_qubits} show a clear trend where increasing the number of qubits in the QCNN architecture enhances the model's accuracy. In this context, accuracy is defined as the proportion of correctly identified events relative to the total number of events in the dataset. This improvement can be attributed to the larger Hilbert space enabling the network to capture more complex patterns and correlations within the data. As demonstrated, the model with 6 qubits at best reaches 90\% and 87. 7\% accuracy in the training and test phase, while the same model with 12 qubits can detect the GRB signal with 99. 38\% and 97. 5\% accuracy in the training and test dataset. However, this gain in accuracy comes at the cost of an increased training time, as the computational complexity increases with the number of qubits, as the 6 qubit model performs faster with a factor of \~ 33. Conversely, reducing the number of qubits below a critical threshold negatively impacts the model's performance, as it becomes insufficient to encode the underlying structure of the data effectively. Therefore, selecting an optimal number of qubits is crucial to balance accuracy and computational efficiency, ensuring the best trade-off between model performance and resource requirements.

\begin{table*}[ht]
\centering
\setlength{\arrayrulewidth}{0.3mm}
\setlength{\tabcolsep}{5pt}
\renewcommand{\arraystretch}{1.5}
\arrayrulecolor{white} 
\setlength{\doublerulesep}{0pt}

\begin{tabular}{|>{\columncolor{tablepurple}}l|                
>{\columncolor{tablepurple}}c|>{\columncolor{tablepurple}}c|>{\columncolor{tablepurple}}c|>{\columncolor{tablepurple}}c|>{\columncolor{tablepurple}}c|}
\hline
\rowcolor{tablepurple!80}
\textbf{NN Model} & \textbf{Num of Qubits} & \textbf{Num of Parameters} & \textbf{Accuracy on Train Set} & \textbf{Accuracy on Test Set} \\ 
\hline
Classical CNN & - & 56 & 99.7\% & 97.35\% \\
\hline
Quantum CNN & 4 & 8 & Doesn’t learn & - \\
\hline
 & 6 & 12 & 90.3\% & 87.7\%  \\
\hline
 & 12 & 24 & 99.31\% & 97.5\% \\
\hline
\end{tabular}
\caption{The effect of number of Qubits on the Performance of QCNN with data reuploading method and its comparison with classical CNN in case of having 440 simulated light curves as training dataset.}
    \label{tab:Num_qubits}
\end{table*}

\subsection{The data encoding effect}
A performance comparison of the QCNN model using amplitude encoding and the data re-uploading method was conducted, Table \ref{tab:data_encoding}. The results indicate that the accuracy of the network drops to 76\% when employing amplitude encoding. This highlights the advantage of the data re-uploading method, which enhances the expressiveness of the variational quantum circuit by iteratively embedding input data at multiple stages of the circuit. This iterative embedding allows the circuit to perform more complex and nonlinear transformations on the input data, making it more capable of capturing intricate patterns and relationships inherent in the dataset. This increased representational power makes the data re-uploading method particularly well-suited for machine learning tasks that require modeling high-dimensional or complex data distributions.

\begin{table*}[ht]
\centering
\setlength{\arrayrulewidth}{0.3mm}
\setlength{\tabcolsep}{5pt}
\renewcommand{\arraystretch}{1.5}

\begin{tabular}{|>{\columncolor{tablepurple}}l|>{\columncolor{tablepurple}}c|>{\columncolor{tablepurple}}c|>{\columncolor{tablepurple}}c|>{\columncolor{tablepurple}}c|>{\columncolor{tablepurple}}c|}
\hline
\rowcolor{tablepurple!80}
\textbf{Data encoding} & \textbf{Num of Qubits} & \textbf{Num of Parameters} & \textbf{Accuracy on Train Set} & \textbf{Accuracy on Test Set} \\ 
\hline
Amplitude & 7 & 14 & 76.56\% & 66.67\%  \\ 
\hline
Data Reuploding & 7 & 63 & 91.13\% & 88.12\%  \\ 
\hline
\end{tabular}

\caption{Performance comparison of QCNN and classical CNN by having Amplitude encoding and data Reuploding method as the encoding method.}
\label{tab:data_encoding}
\end{table*}

\subsection{Few Training Samples}
In the present context, the QCNN has achieved accuracy comparable to that of classical CNNs. To establish a clear quantum advantage, this study focuses on \textit{sample complexity}, which refers to the ability of a model to perform a given task, such as detecting GRB signals, with significantly fewer labeled training samples.
Quantum algorithms are theoretically expected to exhibit superior generalization capabilities compared to classical counterparts \cite{Cerezo:2022nvi}. This advantage arises from their ability to leverage quantum phenomena like superposition and entanglement to encode and process information more efficiently, allowing them to identify underlying patterns and relationships within datasets with fewer data samples \cite{Caro:2021mgf}. Therefore, this study evaluates the QCNN’s performance under conditions of limited training data to explore this potential.

The results in Table \ref{tab:few_training} demonstrate the remarkable generalization capability of the QCNN model when trained on a limited dataset. Specifically, the QCNN achieves high accuracy (95\%) using only 20 light curves in the training set and 180 in the test set. In contrast, the classical CNN fails to reliably detect GRB signals under the same conditions due to the insufficient size of the training dataset. This highlights the strength of QML approaches in scenarios with limited data availability. The observed performance of the QCNN reflects its inherent advantage in sample complexity, showcasing a form of quantum advantage where fewer training samples suffice to achieve superior detection capabilities compared to classical methods. This result is particularly valuable in astrophysical detection tasks, where the rarity of phenomena or observational limitations often result in very few labeled data points for training.

\begin{table*}[ht]
\centering
\setlength{\arrayrulewidth}{0.3mm}
\setlength{\tabcolsep}{5pt}
\renewcommand{\arraystretch}{1.5}

\begin{tabular}{|>{\columncolor{tablepurple}}l|                
>{\columncolor{tablepurple}}c|>{\columncolor{tablepurple}}c|>{\columncolor{tablepurple}}c|>{\columncolor{tablepurple}}c|>{\columncolor{tablepurple}}c|}
\hline
\rowcolor{tablepurple!80}
\textbf{NN Model} & \textbf{Num of Qubits} & \textbf{Num of Parameters} & \textbf{Accuracy on Train Set} & \textbf{Accuracy on Test Set} \\ 
\hline
Classical CNN & - & 56 & 55\% & 52.22\% \\ 
\hline
Quantum CNN & 6 & 24 & 95\% & 98.33\% \\ 
\hline
\end{tabular}

\caption{Performance comparison of QCNN with data reuploading and classical CNN in case of 20 simulated light curves as training dataset.}
\label{tab:few_training}
\end{table*}

\subsection{Performance on Real Data}

To evaluate the robustness of the implemented QCNN beyond simulated datasets, we tested its performance on real observational data from the AGILE satellite. This dataset comprises 43 GRB samples and 101 background samples, presenting significant class imbalance and reflecting the challenges of real-world astrophysical data analysis. This dataset is divided to the training, test and validation set with the rate of 70\%, 20\% and 10\%. For details on the preparation and composition of the AGILE dataset, readers are referred to our previous work \cite{Rizzo:2024wse}. 
As shown in Table \ref{tab:compare_Rizzo} the QCNN demonstrated strong performance on this dataset, achieving 91\% accuracy on the training set and 90\% accuracy on the test set, underscoring its ability to generalize effectively to real data. Also in this case we can observe that Classical CNN reaches very high accuracy (96\% for the training and 93\% for the test set) in detecting the GRB signal. These results indicate that the QCNN is not only capable of detecting GRB signals in simulated datasets but also performs well in real observational scenarios, further highlighting its potential for astrophysical applications.

\begin{table*}[ht]
\centering
\setlength{\arrayrulewidth}{0.3mm}
\setlength{\tabcolsep}{5pt}
\renewcommand{\arraystretch}{1.5}

\begin{tabular}{|>{\columncolor{tablepurple}}l|>{\columncolor{tablepurple}}c|>{\columncolor{tablepurple}}c|>{\columncolor{tablepurple}}c|>{\columncolor{tablepurple}}c|>{\columncolor{tablepurple}}c|}
\hline
\rowcolor{tablepurple!80}
\textbf{NN model} & \textbf{Num of Qubits} & \textbf{Num of Parameters} & \textbf{Accuracy on Train Set} & \textbf{Accuracy on Test Set} \\ 
\hline
Classical CNN & - & 56 & 96.55\% & 93.91\% \\ 
\hline
Quantum CNN  & 7 & 14 & 91.17\% & 90.72\% \\ 
\hline
\end{tabular}

\caption{Performance comparison of QCNN with data reuploading and classical CNN on AGILE dataset.}
\label{tab:compare_Rizzo}
\end{table*}

\section{Conclusion and future work}  
This study represents one of the pioneering efforts to implement QCNN for astrophysical data analysis, specifically for the detection of GRBs. In this context, the detection of GRBs refers to performing a binary classification task using machine learning methods to distinguish GRB signals from background noise.
Different QCNN architectures and various data encoding methods were tested to systematically explore the network's performance. The results demonstrate that QCNNs achieve comparable or superior accuracy to classical CNNs in specific scenarios while requiring fewer parameters, highlighting their efficiency and the potential of QML algorithms.  

A key finding is the generalization power of QCNNs, particularly in cases with limited training data. With only 20 light curves in the training set, the QCNN outperformed classical CNNs, showcasing quantum advantage in terms of sample complexity. Furthermore, the study illustrates the robustness of QCNNs in handling diverse configurations and datasets, including unbalanced, real-world data from the AGILE satellite.  

Despite these promising results, QCNNs face limitations, particularly when implemented on real quantum hardware. The increasing circuit complexity required for larger QCNN architectures can exacerbate the effects of noise and hardware imperfections, potentially degrading model performance. Moreover, the depth of parameterized circuits must be carefully managed to balance representational power and noise resilience. These challenges highlight the need for hardware-aware circuit design and error mitigation techniques to unlock the full potential of QCNNs.

Future research will aim to build on these findings by running the analysis on real quantum hardware to further validate the QCNN's performance. Additionally, efforts will focus on simulating signals with greater complexity, bringing them closer to real GRB signals and backgrounds to better evaluate the QCNN’s capabilities in realistic astrophysical scenarios. This work lays the groundwork for the future integration of quantum computing techniques in astrophysics.  

\section*{Acknowledgment}
This research was supported by the ICSC Italian national research center on High Performance Computing, Big Data, and Quantum Computing. Moreover, this work was conducted in the context of the CTAO consortium.


\begin{thebibliography}{10}

\bibitem{365700}
P.W. Shor.
\newblock Algorithms for quantum computation: discrete logarithms and factoring.
\newblock In {\em Proceedings 35th Annual Symposium on Foundations of Computer Science}, pages 124--134, 1994.

\bibitem{Lloyd1996UniversalQS}
Seth Lloyd.
\newblock Universal quantum simulators.
\newblock {\em Science}, 273:1073 -- 1078, 1996.

\bibitem{Harrow_2009}
Aram~W. Harrow, Avinatan Hassidim, and Seth Lloyd.
\newblock Quantum algorithm for linear systems of equations.
\newblock {\em Physical Review Letters}, 103(15), October 2009.

\bibitem{Zeguendry:2023pbk}
Amine Zeguendry, Zahi Jarir, and Mohamed Quafafou.
\newblock {Quantum Machine Learning: A Review and Case Studies}.
\newblock {\em Entropy}, 25(2):287, 2023.

\bibitem{2019NatPh..15.1273C}
Iris {Cong}, Soonwon {Choi}, and Mikhail~D. {Lukin}.
\newblock {Quantum convolutional neural networks}.
\newblock {\em Nature Physics}, 15(12):1273--1278, December 2019.

\bibitem{Guan:2020bdl}
Wen Guan, Gabriel Perdue, Arthur Pesah, Maria Schuld, Koji Terashi, Sofia Vallecorsa, and Jean-Roch Vlimant.
\newblock {Quantum Machine Learning in High Energy Physics}.
\newblock {\em Mach. Learn. Sci. Tech.}, 2:011003, 2021.

\bibitem{Hassanshahi:2023xgv}
Mohammad~Hassan Hassanshahi, Marcin Jastrzebski, Sarah Malik, and Ofer Lahav.
\newblock {A quantum-enhanced support vector machine for galaxy classification}.
\newblock 6 2023.

\bibitem{Kordzanganeh:2021azm}
Mohammad Kordzanganeh, Aydin Utting, and Anna Scaife.
\newblock {Quantum Machine Learning for Radio Astronomy}.
\newblock In {\em {35th Conference on Neural Information Processing Systems}}, 12 2021.

\bibitem{Cerezo:2022nvi}
M.~Cerezo, Guillaume Verdon, Hsin-Yuan Huang, Lukasz Cincio, and Patrick~J. Coles.
\newblock {Challenges and Opportunities in Quantum Machine Learning}.
\newblock {\em Science}, 2:567--576, 2022.

\bibitem{Bowles:2024fvp}
Joseph Bowles, Shahnawaz Ahmed, and Maria Schuld.
\newblock {Better than classical? The subtle art of benchmarking quantum machine learning models}.
\newblock 3 2024.

\bibitem{Parmiggiani:2023kum}
N.~Parmiggiani et~al.
\newblock {Preliminary Results of a New Deep Learning Method to Detect and Localize GRBs in the AGILE/GRID Sky Maps}.
\newblock In {\em {32th Astronomical Data Analysis Software and Systems}}, 2 2023.

\bibitem{Parmiggiani_2024}
N.~Parmiggiani, A.~Bulgarelli, L.~Castaldini, A.~De Rosa, A.~Di Piano, R.~Falco, V.~Fioretti, A.~Macaluso, G.~Panebianco, A.~Ursi, C.~Pittori, M.~Tavani, and D.~Beneventano.
\newblock A new deep learning model to detect gamma-ray bursts in the agile anticoincidence system.
\newblock {\em The Astrophysical Journal}, 973(1):63, sep 2024.

\bibitem{2023ApJ...945..106P}
N.~{Parmiggiani}, A.~{Bulgarelli}, A.~{Ursi}, A.~{Macaluso}, A.~{Di Piano}, V.~{Fioretti}, A.~{Aboudan}, L.~{Baroncelli}, A.~{Addis}, M.~{Tavani}, and C.~{Pittori}.
\newblock {A Deep-learning Anomaly-detection Method to Identify Gamma-Ray Bursts in the Ratemeters of the AGILE Anticoincidence System}.
\newblock {\em The Astrophysical Journal}{The Astrophysical Journal}, 945(2):106, March 2023.

\bibitem{AGILE:2008nyq}
M.~Tavani et~al.
\newblock {The AGILE Mission}.
\newblock {\em Astron. Astrophys.}, 502:995--1013, 2009.

\bibitem{Rizzo:2024wse}
A.~Rizzo et~al.
\newblock {Quantum Convolutional Neural Networks for the detection of Gamma-Ray Bursts in the AGILE space mission data}.
\newblock In {\em {33th Astronomical Data Analysis Software and Systems}}, 4 2024.

\bibitem{Bergholm:2018cyq}
Ville Bergholm et~al.
\newblock {PennyLane: Automatic differentiation of hybrid quantum-classical computations}.
\newblock 11 2018.

\bibitem{Crooks:2019zff}
Gavin~E. Crooks.
\newblock {Gradients of parameterized quantum gates using the parameter-shift rule and gate decomposition}.
\newblock 5 2019.

\bibitem{Mitarai_2018}
K.~Mitarai, M.~Negoro, M.~Kitagawa, and K.~Fujii.
\newblock Quantum circuit learning.
\newblock {\em Physical Review A}, 98(3), September 2018.

\bibitem{Cerezo_2021}
M.~Cerezo, Andrew Arrasmith, Ryan Babbush, Simon~C. Benjamin, Suguru Endo, Keisuke Fujii, Jarrod~R. McClean, Kosuke Mitarai, Xiao Yuan, Lukasz Cincio, and Patrick~J. Coles.
\newblock Variational quantum algorithms.
\newblock {\em Nature Reviews Physics}, 3(9):625–644, August 2021.

\bibitem{P_rez_Salinas_2020}
Adrián Pérez-Salinas, Alba Cervera-Lierta, Elies Gil-Fuster, and José~I. Latorre.
\newblock Data re-uploading for a universal quantum classifier.
\newblock {\em Quantum}, 4:226, February 2020.

\bibitem{2009A&A...496..585G}
G.~{Ghirlanda}, L.~{Nava}, G.~{Ghisellini}, A.~{Celotti}, and C.~{Firmani}.
\newblock {Short versus long gamma-ray bursts: spectra, energetics, and luminosities}.
\newblock {\em Astronomy \& Astrophysics}, 496(3):585--595, March 2009.

\bibitem{Fynbo:2006du}
J.~P.~U. Fynbo et~al.
\newblock {Probing cosmic chemical evolution with gamma-ray bursts: grb060206 at z=4.048}.
\newblock {\em Astron. Astrophys.}, 451:L47--L50, 2006.

\bibitem{2009astro2010S.284S}
M.~{Stamatikos}, N.~{Gehrels}, F.~{Halzen}, P.~{M{\'e}sz{\'a}ros}, and P.~W.~A. {Roming}.
\newblock {Multi-Messenger Astronomy with GRBs: A White Paper for the Astro2010 Decadal Survey}.
\newblock In {\em astro2010: The Astronomy and Astrophysics Decadal Survey}, volume 2010, page 284, January 2009.

\bibitem{Thompson:2022ufx}
David~J. Thompson and Colleen~A. Wilson-Hodge.
\newblock {Fermi Gamma-ray Space Telescope}.
\newblock 10 2022.

\bibitem{2020ApJ...893...46V}
A.~{von Kienlin}, C.~A. {Meegan}, W.~S. {Paciesas}, P.~N. {Bhat}, E.~{Bissaldi}, M.~S. {Briggs}, E.~{Burns}, W.~H. {Cleveland}, M.~H. {Gibby}, M.~M. {Giles}, A.~{Goldstein}, R.~{Hamburg}, C.~M. {Hui}, D.~{Kocevski}, B.~{Mailyan}, C.~{Malacaria}, S.~{Poolakkil}, R.~D. {Preece}, O.~J. {Roberts}, P.~{Veres}, and C.~A. {Wilson-Hodge}.
\newblock {The Fourth Fermi-GBM Gamma-Ray Burst Catalog: A Decade of Data}.
\newblock {\em The Astrophysical Journal}, 893(1):46, April 2020.

\bibitem{2021ApJ...913...60P}
S.~{Poolakkil}, R.~{Preece}, C.~{Fletcher}, A.~{Goldstein}, P.~N. {Bhat}, E.~{Bissaldi}, M.~S. {Briggs}, E.~{Burns}, W.~H. {Cleveland}, M.~M. {Giles}, C.~M. {Hui}, D.~{Kocevski}, S.~{Lesage}, B.~{Mailyan}, C.~{Malacaria}, W.~S. {Paciesas}, O.~J. {Roberts}, P.~{Veres}, A.~{von Kienlin}, and C.~A. {Wilson-Hodge}.
\newblock {The Fermi-GBM Gamma-Ray Burst Spectral Catalog: 10 yr of Data}.
\newblock {\em The Astrophysical Journal}, 913(1):60, May 2021.

\bibitem{SwiftScience:2004ykd}
N.~Gehrels et~al.
\newblock {The Swift Gamma-Ray Burst Mission}.
\newblock {\em Astrophys. J.}, 611:1005--1020, 2004.
\newblock [Erratum: Astrophys.J. 621, 558 (2005)].

\bibitem{2022ApJ...925..152U}
A.~{Ursi}, M.~{Romani}, F.~{Verrecchia}, C.~{Pittori}, M.~{Tavani}, M.~{Marisaldi}, M.~{Galli}, C.~{Labanti}, N.~{Parmiggiani}, A.~{Bulgarelli}, A.~{Addis}, L.~{Baroncelli}, M.~{Cardillo}, C.~{Casentini}, P.~W. {Cattaneo}, A.~{Chen}, A.~{Di Piano}, F.~{Fuschino}, F.~{Longo}, F.~{Lucarelli}, A.~{Morselli}, G.~{Piano}, and S.~{Vercellone}.
\newblock {The Second AGILE MCAL Gamma-Ray Burst Catalog: 13 yr of Observations}.
\newblock {\em The Astrophysical Journal}, 925(2):152, February 2022.

\bibitem{CTAConsortium:2017dvg}
B.~S. Acharya et~al.
\newblock {\em {Science with the Cherenkov Telescope Array}}.
\newblock WSP, 11 2018.

\bibitem{doi:10.1142/10986}
{\em Science with the Cherenkov Telescope Array}.
\newblock WORLD SCIENTIFIC, 2019.

\bibitem{gammapy:2023}
Axel {Donath}, R\'egis {Terrier}, Quentin {Remy}, Atreyee {Sinha}, Cosimo {Nigro}, Fabio {Pintore}, Bruno {Kh\'elifi}, Laura {Olivera-Nieto}, Jose~Enrique {Ruiz}, Kai {Br\"ugge}, Maximilian {Linhoff}, Jose~Luis {Contreras}, Fabio {Acero}, Arnau {Aguasca-Cabot}, David {Berge}, Pooja {Bhattacharjee}, Johannes {Buchner}, Catherine {Boisson}, David {Carreto Fidalgo}, Andrew {Chen}, Mathieu {de Bony de Lavergne}, Jos\'e~Vinicius {de Miranda Cardoso}, Christoph {Deil}, Matthias {F\"u\ss{}ling}, Stefan {Funk}, Luca {Giunti}, Jim {Hinton}, L\'ea {Jouvin}, Johannes {King}, Julien {Lefaucheur}, Marianne {Lemoine-Goumard}, Jean-Philippe {Lenain}, Rub\'en {L\'opez-Coto}, Lars {Mohrmann}, Daniel {Morcuende}, Sebastian {Panny}, Maxime {Regeard}, Lab {Saha}, Hubert {Siejkowski}, Aneta {Siemiginowska}, Brigitta~M. {Sip"ocz}, Tim {Unbehaun}, Christopher {van Eldik}, Thomas {Vuillaume}, and Roberta {Zanin}.
\newblock Gammapy: A python package for gamma-ray astronomy.
\newblock {\em A\&A}, 678:A157, 2023.

\bibitem{universe7100374}
Cosimo Nigro, Tarek Hassan, and Laura Olivera-Nieto.
\newblock Evolution of data formats in very-high-energy gamma-ray astronomy.
\newblock {\em Universe}, 7(10), 2021.

\bibitem{gammapy:2024}
Fabio Acero, Juan Bernete, Noah Biederbeck, Julia Djuvsland, Axel Donath, Kirsty Feijen, Stefan Fröse, Claudio Galelli, Bruno Khélifi, Jana Konrad, Paula Kornecki, Maximilian Linhoff, Kurt McKee, Simone Mender, Daniel Morcuende, Laura Olivera-Nieto, Fabio Pintore, Michael Punch, Maxime Regeard, Quentin Remy, Atreyee Sinha, Hanna Stapel, Katrin Streil, Régis Terrier, and Tim Unbehaun.
\newblock Gammapy: Python toolbox for gamma-ray astronomy, https://doi.org/10.5281/zenodo.10726484, 02 2024.

\bibitem{2022icrc.confE.998P}
B.~{Patricelli}, A.~{Carosi}, L.~{Nava}, M.~{Seglar-Arroyo}, F.~{Sch{\"u}ssler}, A.~{Stamerra}, A.~{Adelfio}, H.~{Ashkar}, A.~{Bulgarelli}, T.~{Di Girolamo}, A.~{Di Piano}, T.~{Gasparetto}, J.~G. {Green}, F.~{Longo}, I.~{Agudo}, A.~{Berti}, E.~{Bissaldi}, G.~{Cella}, A.~{Circiello}, S.~{Covino}, G.~{Ghirlanda}, B.~{Humensky}, S.~{Inoue}, J.~{Lefaucheur}, M.~D. {Filipovic}, M.~{Razzano}, D.~{Ribeiro}, O.~{Sergijenko}, G.~{Stratta}, and S.~{Vergani}.
\newblock {Searching for very-high-energy electromagnetic counterparts to gravitational-wave events with the Cherenkov Telescope Array}.
\newblock In {\em 37th International Cosmic Ray Conference}, page 998, March 2022.

\bibitem{cherenkov_telescope_array_observatory_2021_5499840}
Cherenkov Telescope~Array Observatory and Cherenkov Telescope~Array Consortium.
\newblock {CTAO Instrument Response Functions - prod5 version v0.1, https://doi.org/10.5281/zenodo.5499840}, September 2021.

\bibitem{Sim:2019yyv}
Sukin Sim, Peter~D. Johnson, and Al\'an Aspuru-Guzik.
\newblock {Expressibility and Entangling Capability of Parameterized Quantum Circuits for Hybrid Quantum-Classical Algorithms}.
\newblock {\em Adv. Quantum Technol.}, 2(12):1900070, 2019.

\bibitem{Caro:2021mgf}
Matthias~C. Caro, Hsin-Yuan Huang, M.~Cerezo, Kunal Sharma, Andrew Sornborger, Lukasz Cincio, and Patrick~J. Coles.
\newblock {Generalization in quantum machine learning from few training data}.
\newblock {\em Nature Commun.}, 13(1):4919, 2022.

\end{thebibliography}
\end{document}